\documentclass[prd,showpacs,preprintnumbers,twocolumn,amsmath,nofootinbib,amssymb]{revtex4}
\usepackage{graphicx,color,dcolumn,booktabs,bm}
\usepackage{epstopdf}
\usepackage{longtable,lscape}
\usepackage{txfonts}
\usepackage{overpic}
\usepackage{diagbox}
\usepackage{slashbox}
\usepackage{float}
\usepackage{amssymb}
\usepackage{epstopdf}
\usepackage{indentfirst}
\usepackage{feynmf}   
\usepackage{slashed}  
\usepackage{cases}
\usepackage{ulem}
\usepackage{color}
\usepackage{float}
\usepackage{multirow}
\usepackage{tikz}
\usepackage{epstopdf}
\usepackage{epsfig,dsfont,amssymb,amsmath,amsfonts,amsbsy,mathrsfs}
\usepackage{multirow}
\usepackage{graphicx}  
\usepackage{float}  
\usepackage{subfigure}  

\usepackage[colorlinks, citecolor=blue,anchorcolor=red,menucolor=red, linkcolor=red,filecolor=red,runcolor=red,urlcolor=blue,frenchlinks=red]{hyperref}
\makeatletter
\@addtoreset{equation}{section}
\makeatother

\newcommand{\nc}{\newcommand}
\nc{\tj}[1]{\textcolor{red}{Tianjie: #1}}
\allowdisplaybreaks

\begin{document}
\title{Can the three new states around 2.2 GeV assign to  $\omega(3D)$  }
\author{Ya-Rong Wang$^{1,2}$}\email{nanoshine@foxmail.com}
\author{Yan Ma$^{1,2}$}\email{mycurry302022@163.com}
\author{Cheng-Qun Pang$^{1,2}$\footnote{Corresponding author}}\email{pcq@qhnu.edu.cn}
\affiliation{$^1$College of Physics and Electronic Information Engineering, Qinghai Normal University, Xining 810000, China\\$^2$Joint Research Center for Physics,
Lanzhou University and Qinghai Normal University,
Xining 810000, China }

\begin{abstract}
Recently, the BESIII Collaboration reported three resonances: $X(2232)$ with $M = 2232 \pm 19 \pm 27$ MeV and  $\Gamma = 93 \pm 53 \pm 20$ MeV, $X(2200)$ whose mass $M = 2200 \pm 11 \pm 17$ MeV and width $\Gamma = 74 \pm 20 \pm 24$ MeV as well as $X(2222)$ which has mass of $2222 \pm 7 \pm 2$ MeV and the width of $59 \pm 30 \pm 6$ MeV. The mass spectrum of $\omega$ meson family is studied utilizing the modified Godfrey-Isgur model, and the two-body strong decays of $X(2232)$, $X(2200)$ and $X(2222)$ within two different approaches of the $^3P_0$ model. We find that the newly discovered states $X(2232)$, $X(2200)$ and $X(2222)$ may be the same and are most likely to be the $\omega(3D)$ state. The discovery could be useful in establishing entire $\omega$ mesons.     

\end{abstract}
\pacs{14.40.Be, 12.38.Lg, 13.25.Jx}
\maketitle
\section{Introduction}
Light flavor physics is a fascinating and significant research field. In the past few decades, more and more novel hadronic states have been discovered due to the improvement of particle detector accuracy.
Very recently, the BESIII Collaboration has announced the discovery of two new states (X(2232) and X(2200)) in the measurement of the $e^+e^- \to \omega\pi^+\pi^-$ cross section at $\sqrt{s} = 2.00 \to 3.080$ GeV \cite{BESIII:2022yzp}. 
$X(2232)$ has the mass of $2232 \pm 19 \pm 27$ MeV and width of $93 \pm 53 \pm 20$ MeV, according to the measurement. 
$X(2200)$ is discovered with a combined significance of $7.9\sigma$, and its mass and width were determined to be $M = 2200 \pm 11 \pm 17$ MeV and $\Gamma = 74 \pm 20 \pm 24$ MeV, respectively. The BESIII Collaboration also announced the observation of the $X(2222)$ state in the $e ^+e^- \to \omega\pi^0\pi^0$ process \cite{BESIII:2021uni}, which has a statistical significance larger than $5\sigma$. $X(2220)$ has the mass of $M = 2222 \pm 7 \pm 2$ MeV and the width of $\Gamma = 59 \pm 30 \pm 6$ MeV, according to the measurements. Additionally, $X(2222)$ may be an $\omega(3D)$ state, according to our earlier work \cite{Wang:2022xxi},  Lanzhou group  indicates that the enhancement structures around 2.2 GeV existing in $e^+e^- \to \omega\eta$ and $e^+e^- \to \omega\pi^0\pi^0$ contain the $\omega(4S)$ and $\omega(3D)$ signals \cite{Zhou:2022wwk}. The BESIII Collaboration announced $X(2222)$ could be an excited $\omega$ resonances \cite{BESIII:2022mxl}.

In 2002, $\omega(2205)$ state was firstly observed in the $\bar{p}p \to \omega\eta$ and $\omega\pi^0\pi^0$ process with $M = 2205 \pm 30$ MeV and  $\Gamma = 350 \pm 90$ MeV  \cite{Anisovich_2002}. 
According to the authors, this state is most likely the radial excitation of $\omega(1650)$ and it couples strongly to $^3D_1$ \cite{Anisovich_2002}. Furthermore, they also stated that this resonance has quite a large $^3D_1$ amplitude, and is consistent with the $^3D_1$ state expected around 2230 MeV.  $X(2232)$, $X(2200)$ and $X(2222)$ states have adjacent masses and their widths are relatively narrow. This phenomenon is so intriguing that it catches our attention. According to our previous work, there is a great possibility that these three states can be grouped into $\omega$ meson family.  $\omega$ meson plays a vital role in the light meson families. This work helps us reveal the inner structure of $\omega(3D)$, which is crucial to understanding the $\omega$ meson family.  
\par
In this work, we shall adopt the modified Godfrey-Isgur (MGI) model to estimate the mass spectrum of the $\omega$ states and calculate the OZI-allowed two-body strong decays of $X(2232)$, $X(2200)$ and $X(2222)$ within two different approaches of the $^3P_0$ model. Our results show that the enhancement structures $X(2232)$, $X(2200)$ and $X(2222)$ may be the same states, they are all the $\omega(3D)$ states.   

\section{Mass spectrum and Okubo-Zweig-Iizuka allowed two-body strong decay}

\subsection{A brief review of MGI model}

In 1985, Godfrey and Isgur proposed the GI model for describing relativistic meson spectra with great success \cite{Godfrey:1985xj},  Q.-T. Song $et~al$. \cite{Song:2015nia,Song:2015fha} introduced the color screening effect on the basis of the Cornell potential to modify the GI model (MGI model).

The basic equation of the model is the rest frame Schr\"{o}dinger-type equation \cite{Godfrey:1985xj}:
\begin{equation}
H|\psi\rangle = E|\psi\rangle,
\end{equation}
the Hamiltonian of the potential model and can be written as
\begin{equation}\label{Hamtn}
  \tilde{H}=\sqrt{m_1^2+\mathbf{p}^2}+\sqrt{m_2^2+\mathbf{p}^2}+\tilde{V}_{\mathrm{eff}}(\mathbf{p,r}),
\end{equation}
where $m_1$ and $m_2$ denote the masses of quarks and antiquarks, respectively, the
 effective potential has a familiar format in the nonrelativistic limit \cite{Godfrey:1985xj,Lucha:1991vn}
\begin{eqnarray}
V_{\mathrm{eff}}(r)=H^{\mathrm{hyp}}+H^{\mathrm{conf}}+H^{\mathrm{so}}\label{1},
\end{eqnarray}
with
\begin{align}
H^{\mathrm{hyp}}&=-\frac{\alpha_s(r)}{m_{1}m_{2}}\Bigg[\frac{8\pi}{3}\bm{S}_1\cdot\bm{S}_2\delta^3 (\bm r) +\frac{1}{r^3}\Big(\frac{3\bm{S}_1\cdot\bm r \bm{S}_2\cdot\bm r}{r^2} \nonumber  \\ \label{3.1}
&\quad -\bm{S}_1\cdot\bm{S}_2\Big)\Bigg] (\bm{F}_1\cdot\bm{F}_2), \\
 H^{\mathrm{conf}}&=\Big[-\frac{3}{4}(\frac{b(1-e^{-\mu r})}{\mu}+c)+\frac{\alpha_s(r)}{r}\Big](\bm{F}_1\cdot\bm{F}_2),\\ 
H^{\mathrm{so}}=&H^{\mathrm{so(cm)}}+H^{\mathrm{so(tp)}},  
\end{align}
in color space, $\bm{F}$ are related to the Gell-Mann matrices. $\bm{S}_1/\bm{S}_2$ indicates the spin of quark/antiquark and $\bm{L}$ is the orbital momentum. 
 All the parameters involved in the MGI model are listed in Table \ref{parameter} given by Ref. \cite{Wang:2021abg}, where a systematic study in performed for the spectroscopy behavior of light unflavored vector mesons with mass at the range of $2.4 - 3$ GeV.  More details of the MGI model can be seen in Ref. \cite{Wang:2022xxi}.

\begin{table}[htbp]
\caption{Parameters and their values in this work. \label{parameter}}
\begin{center}
\begin{tabular}{cccc}
\toprule[1pt]\toprule[1pt]
Parameter &  Value \cite{Wang:2021abg}&Parameter &  Value \cite{Wang:2021abg} \\
 \midrule[1pt]
$m_u$ (GeV)       &0.22         &{$\sigma_0$ (GeV)}    &{1.8}\\
$m_d$ (GeV)       &0.22         &{$s$ (GeV)}            &{3.88}\\
$m_s$ (GeV)       &0.424        &$\mu$ (GeV)           &0.081 \\
$b$ (GeV$^2$)     &0.229        &$c$ (GeV)             &-0.3\\
$\epsilon_c$      &-0.164       &$\epsilon_{sov}$      &0.262\\
$\epsilon_{sos}$  &0.9728       & $\epsilon_t$         &1.993\\
\bottomrule[1pt]\bottomrule[1pt]
\end{tabular}
\end{center}
\end{table}


We calculate the mass spectrum of $\omega$ meson family and obtain the masses of $\omega(3S)$, $\omega(4S)$, $\omega(5S)$, $\omega(1D)$, $\omega(2D)$ and $\omega(3D)$ are 1861.6 MeV, 2181.6 MeV, 2429.0 MeV, 2633.1 MeV, 2003.0 MeV and 2284.1 MeV, respectively. 
The new observed states $X(2200)$, $X(2232)$ and $X(2222)$ have approximate masses $2200 \pm 11 \pm 17$ MeV, $2232 \pm 19 \pm 27$ MeV and $2222 \pm 7 \pm 2$ MeV, respectively. According
to their mass spectrum we obtained, these three states may be $\omega(3D)$ states.

\subsection{A brief review of QPC model}

The $^3P_0$ model is widely applied in the OZI-allowed two-body strong decays of hadrons in Refs. \cite{vanBeveren:1979bd, vanBeveren:1982qb, Capstick:1993kb, Page:1995rh, Titov:1995si, Ackleh:1996yt, Blundell:1996as,
Bonnaz:2001aj, Zhou:2004mw, Lu:2006ry, Zhang:2006yj, Luo:2009wu, Sun:2009tg, Liu:2009fe, Sun:2010pg, Rijken:2010zza, Ye:2012gu,
Wang:2012wa, He:2013ttg, Sun:2013qca, Pang:2014laa, Wang:2014sea, Chen:2015iqa, Pang:2017dlw, Pang:2018gcn}.

In our previous treatment of the two-body decays $A \to BC$, the decay widths were obtained from the amplitudes read
\begin{eqnarray}
\Gamma&=&\frac{\pi}{4} \frac{|\mathbf{P}|}{m_A^2}\sum_{J,L}|\mathcal{M}^{JL}(\mathbf{P})|^2,
\end{eqnarray}
where $m_{A}$ is the mass of an initial state $A$, and
the two decay amplitudes can be related by the Jacob-Wick formula \cite{Jacob:1959at} as

\begin{equation}
\begin{aligned}
\mathcal{M}^{JL}(\mathbf{P}) = &\frac{\sqrt{4\pi(2L+1)}}{2J_A+1}\sum_{M_{J_B}M_{J_C}}\langle L0;JM_{J_A}|J_AM_{J_A}\rangle \\
    &\times \langle J_BM_{J_B};J_CM_{J_C}|{J_A}M_{J_A}\rangle \mathcal{M}^{M_{J_{A}}M_{J_B}M_{J_C}}.
\end{aligned}	
\end{equation}

The QPC model's input parameter $\gamma$, has two different values from method A and method B. In method A,  $\gamma$ is taken to be 8.7 for $u \bar u/d \bar d$ pair creation, while the strength of the $s\bar s$ pair creation satisfies $\gamma=8.7/\sqrt{3}$ \cite{LeYaouanc:1977gm,Ye:2012gu,He:2013ttg}. In method B, we adopt $\gamma = 8.1$, which was obtained by fitting the decay width of $\omega(1420)$ \cite{Zyla:2020zbs} state.

We use two methods to verify the conclusion. The parameter $R$ of the final channel in the simple harmonic oscillator wave function is determined by reproducing the root mean square radius by solving the Schr\"odinger equation with the effective potential \cite{Close:2005se}. We may set a $R$ range between the  $R$ values determined by the potential model in Refs. \cite{Close:2005se,Godfrey:1985xj} for these three states for method A. For method B, We use a simple harmonic oscillator  wave function $\psi_{n\ell m}({\bf k})=R_{n\ell}(R,{\bf k})\mathcal{Y}_{n\ell m}(\bf{k})$ to describe the meson wave function involved in the decays when calculating the spatial integral of the decay amplitude.

\begin{table}
\centering
\caption{The total and partial decay widths of the  $X(2222)$, $X(2200)$  and $X(2232)$ when treated them as $\omega(3D)$ states in method B, the unit of  widths is MeV. The $\gamma$ value is 8.1. Exp. means the experiment width.   \label{tabdecay}}
\[\begin{array}{c|c|c|c}
\toprule[1pt]\toprule[1pt]
               &X(2222)  &X(2200) &X(2232) \\\midrule[1pt]
\text{Exp.}            &59 \pm 30.59  &74 \pm 31.24     &93 \pm 56.65\\
\text{Total}    &{73.1}        &{65.0}           &{67.3}  \\
{b_1}\pi        &{35.9}        &{31.2}           &{27.2} \\
\pi\rho(1450)   &{10.6}        &{9.46}           &{9.58} \\
\rho\pi         &{5.90}        &{2.99}           &{4.05} \\
\rho{a_1}       &{4.40}        &{4.66}           &{7.87} \\
\rho{a_2}       &{3.47}        &{3.11}           &{5.41} \\
\eta{h_1}       &{3.02}        &{2.35}           &{1.53}\\
\pi\rho_3       &{1.72}        &{1.47}           &{2.84} \\
\omega{f_2}     &{1.72}        &{1.82}           &{2.84}\\
\omega{f_1}     &{1.13}        &{1.93}           &{2.12}\\
\pi\rho(1700)   &{1.03}        &{0.971}           &{1.16} \\
KK(1460)        &{1.11}       &{0.882}          &{0.000297} \\
KK              &{1.02}       &{0.945}          &{0.312} \\
\eta\omega(1420)&{0.438}       &{0.277}          &{0.202} \\
K{K_1^\prime}   &{0.291}       &{0.0999}          &{0.112} \\
\omega\eta      &{0.364}       &{0.289}          &{0.117} \\
\rho\pi(1300)   &{0.264}       &{0.619}          &{0.0640} \\
KK^*(1410)      &{0.358}       &{0.0490}         &{0.0412} \\
\omega{f_0(1370)}     &0.135         &0.0735           & 0.160        \\
\eta^\prime{h_1}  &{0.0901 }   &{0.178}          &{0.318} \\
\omega\eta(1295)  &{0.0424}    &{0.116}          &{0.0137}\\
K{K_1}          &{0.0607}      &{1.35}           &{1.14} \\
\omega\eta^\prime &{0.00357}   &{0.121}          &{0.103} \\
KK^*            &{0.00342}     &{0.121}          &{0.106} \\
\bottomrule[1pt]\bottomrule[1pt]
\end{array}\]
\end{table}

\subsection{Decay modes and widths}

It should be more emphasis on the fact that the mass alone is insufficient to identify this assignment, and decay behaviors need to be analyzed. We adopt two methods to show the widths of $X(2232)$, $X(2200)$ and $X(2222)$, when they are treated as $\omega(3D)$. According to our results of the QPC model, $X(2232)$, $X(2200)$ and $X(2222)$ can decay into $b_1\pi$, $\pi\rho(1450)$, $\rho{a_1}$, $\eta\omega(1420)$, $KK$, $KK(1460)$, $\omega\eta$, $\pi\rho_3$, $\eta{h_1}$, $\rho\pi$, $\rho{a_2}$, $\omega{f_2}$, $\rho\pi(1300)$, $\omega{f_1}$, $KK_1(1400)$, $KK_1(1270)$, $\pi\rho(1700)$, $KK^*$, $KK^*(1410)$, $\omega\eta^\prime$, $\eta^\prime{h_1}$, $\omega\eta(1295)$ and so on. The decay modes ${b_1}\pi$, $\omega{f_0(1370)}$ and $\omega{f_2}$ are also proved in the experiment \cite{BESIII:2022yzp}.

\subsubsection{Results of method A}

\begin{figure*}[htbp]
\centering%
\includegraphics[height=1.5in, width=7in]{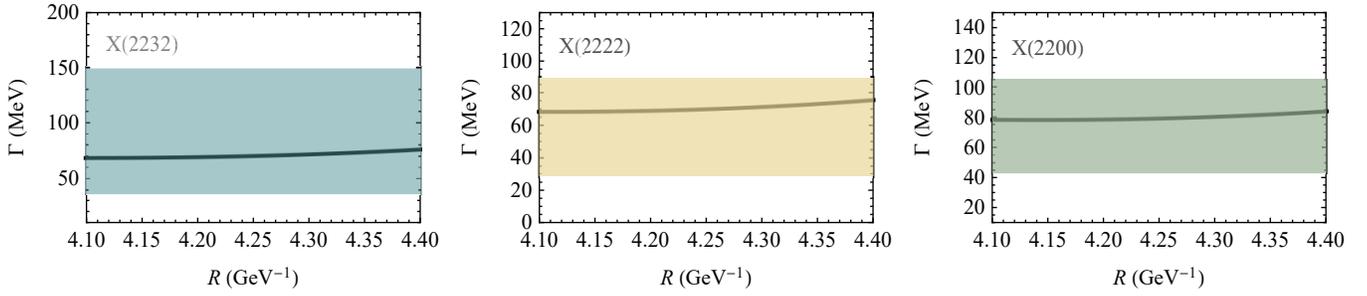}
\caption{The $R$ dependence of the calculated total decay widths of $X(2222)$, $X(2200)$  and $X(2232)$ in method A. The colored band is the experiment width of $X(2222)$, $X(2200)$  and $X(2232)$ from BESIII \cite{2022arXiv220804507B,BESIII:2021uni}. \label{total} }
\end{figure*}

\begin{figure*}[htbp]
\centering%
\includegraphics[height=2.5in, width=6.5in]{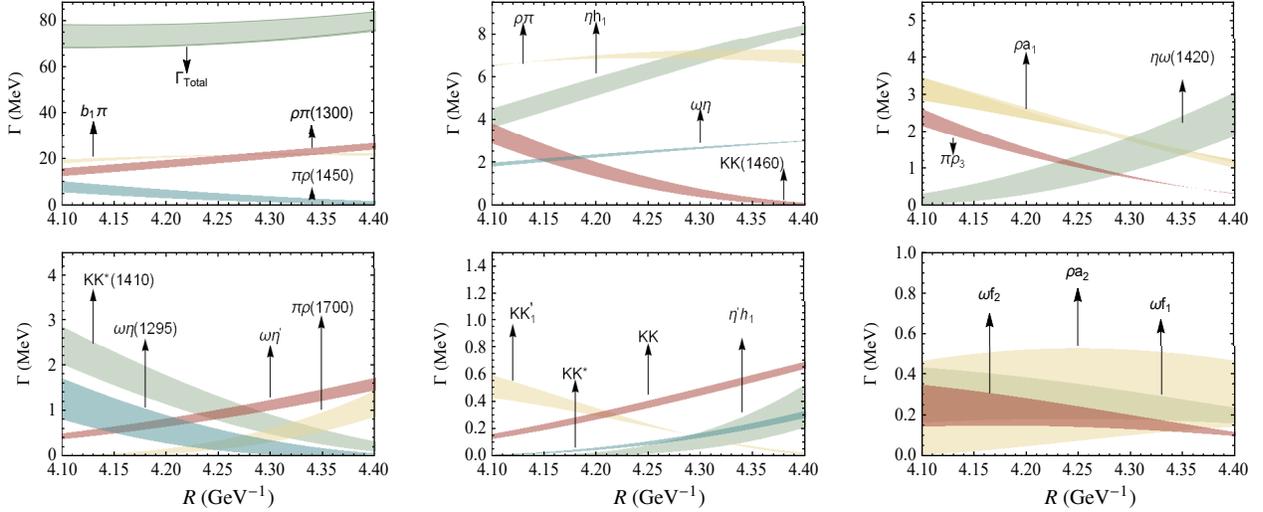}
\put(-77,-10){\footnotesize {$R~(\text{GeV}^{-1})$}}
\put(-240,-10){\footnotesize {$R~(\text{GeV}^{-1})$}}
\put(-403,-10){\footnotesize {$R~(\text{GeV}^{-1})$}}
\caption{The $R$ dependence of the calculated partial and total decay widths of $X(2222)$, $X(2200)$  and $X(2232)$ in method A. Here, we do not list the tiny decay modes. $b_1$, $h_1$,$a_1$, $a_2$, $f_1$, $f_2$, $K_1$ and ${K_1}^\prime$ represent $b_1(1235)$, $h_1(1170)$, $a_1(1260)$, $a_2(1320)$, $f_1(1285)$, $f_2(1270)$, $K_1(1270)$ and ${K_1(1400)}$. \label{decay}}
\end{figure*}

We obtained the $R$ dependence of the calculated partial and total decay widths of $X(2222)$, $X(2200)$ and $X(2232)$ states in the method A, as shown in Fig. \ref{total} and \ref{decay}. 
In Fig. \ref{decay}, partial decay widths of $X(2222)$, $X(2200)$  and $X(2232)$ are calculated by us and the colorful band are the experiment widths of $X(2222)$, $X(2200)$ and $X(2232)$ in Fig. \ref{total}. The main decay channel is ${b_1}\pi$ and it has been seen in the experiment \cite{BESIII:2022yzp}, which also proves our assumption. The total widths we obtained of $X(2222)$, $X(2200)$ and $X(2232)$ are about $68 - 81$ MeV, $78 - 89$ MeV and $68 - 82$ MeV, respectively, which are in good agreement with the BESIII measurement of $59 \pm 30 \pm 6$ MeV, $74 \pm 20 \pm 24$ MeV and $93 \pm 53 \pm 20 $MeV, respectively.    

\subsubsection{Results of method B}

We employ the QPC model with the realistic meson wave functions obtained from the MGI model to evaluate the decay widths of $X(2222)$, $X(2200)$  and $X(2232)$ when they are treated as $\omega(3D)$ states in the method B. The total widths and partial widths of these three states are listed in Table \ref{decay}. The total widths of $X(2222)$, $X(2200)$  and $X(2232)$ are expected to be about $73$ MeV, $65$ MeV and $67$ MeV, respectively, which agrees well with the BESIII measurement of $\Gamma_{X(2222)} = 59 \pm 30.59$ MeV, $\Gamma_{X(2232)} = 93 \pm 56.65$ MeV and $\Gamma_{X(2200)} = 74 \pm 31.24$ MeV. Furthermore, the crucial decay modes of $X(2222)$, $X(2200)$  and $X(2232)$  are $b_1\pi$, $\pi\rho(1450)$, $\rho\pi$ and $\rho{a_1}$. In addition, $\rho{a_2}$, $\eta{h_1}$, $\pi\rho_3$ and $\omega{f_2}$ are the sizeable decay modes as well. Other decay channels, such as $KK^*$, $\omega\eta^\prime$, $K{K_1}$, $\omega\eta(1295)$ and so on, are too narrow for experiment to find. We hope more precise experimental data can be released in the future to reveal the inner structure of $\omega(3D)$ state.

\section{Summary}
In this work, we perform a study on the properties for $X(2232)$, $X(2200)$ and $X(2222)$. The mass spectrum of $\omega$ indicates that these three states may be $\omega(3D)$ states. Additionally, the decay widths are calculated in the QPC model with two methods and the result points out when we treated $X(2232)$, $X(2200)$ and $X(2222)$ as $\omega(3D)$ states, the widths are in good agreement with experiment values. The most important decay mode ${b_1}\pi$ in our theoretical calculation is also proved in the experiment of BESIII Collaboration. Other decay channels like $\omega{f_0(1370)}$ and $\omega{f_2}$ are found both in the theoretical research and experiment. We expect that more and more results will be released by lattice QCD in the near future and further experimental progress on the $\omega(3D)$.

\section{Acknowledgements}
This work is supported  by the National Natural Science Foundation of China under Grants No. 11965016 and Department of Qinghai Province  No. 2020-ZJ-728.

\bibliographystyle{apsrev4-1}
\bibliography{hepref}
\end{document}